\documentclass[conference]{IEEEtran}
\IEEEoverridecommandlockouts
\usepackage{cite}
\usepackage{amsmath,amssymb,amsfonts}
\usepackage{algorithmic}
\usepackage{graphicx}
\usepackage{textcomp}
\usepackage{xcolor}
\usepackage{longtable}
\usepackage{listings}
\usepackage{hyperref}
\usepackage{makecell}
\usepackage{multirow}
\usepackage{tikz}
\def\BibTeX{{\rm B\kern-.05em{\sc i\kern-.025em b}\kern-.08em
    T\kern-.1667em\lower.7ex\hbox{E}\kern-.125emX}}
    
\newcommand\copyrighttext{%
  \footnotesize \textcopyright 2022 IEEE. Personal use of this material is permitted. Permission from IEEE must be obtained for all other uses, in any current or future media, including reprinting/republishing this material for advertising or promotional purposes, creating new collective works, for resale or redistribution to servers or lists, or reuse of any copyrighted component of this work in other works.
  DOI: 10.1109/CIW-IUS56691.2022.00007}
\newcommand\copyrightnotice{%
\begin{tikzpicture}[remember picture,overlay]
\node[anchor=south,yshift=10pt] at (current page.south) {\fbox{\parbox{\dimexpr\textwidth-\fboxsep-\fboxrule\relax}{\copyrighttext}}};
\end{tikzpicture}%
}
    
\begin{document}

\title{Web-based volunteer distributed computing for handling time-critical urgent workloads}

\author{\IEEEauthorblockN{Nick Brown}
\IEEEauthorblockA{\textit{EPCC} \\
\textit{The University of Edinburgh}\\
The Bayes Centre, Edinburgh, UK \\
n.brown@epcc.ed.ac.uk}
\and
\IEEEauthorblockN{Simon Newby}
\IEEEauthorblockA{\textit{EPCC} \\
\textit{The University of Edinburgh}\\
The Bayes Centre, Edinburgh, UK}
}

\maketitle
\copyrightnotice

\begin{abstract}
Urgent computing workloads are time critical, unpredictable, and highly dynamic. Whilst efforts are on-going to run these on traditional HPC machines, another option is to leverage the computing power donated by volunteers. Volunteer computing, where members of the public donate some of their CPU time to large scale projects has been popular for many years because it is a powerful way of delivering compute for specific problems, with the public often eager to contribute to a good cause with societal benefits. However, traditional volunteer computing has required user installation of specialist software which is a barrier to entry, and the development of the software itself by the projects, even on-top of existing frameworks, is non-trivial. 

As such, the number of users donating CPU time to these volunteer computing projects has decreased in recent years, and this comes at a time when the frequency of disasters, often driven by climate change, are rising fast. We believe that an alternative approach, where visitors to websites donate some of their CPU time whilst they are browsing, has the potential to address these issues. However, web-based distributed computing is an immature field and there are numerous questions that must be answered to fully understand the viability of leveraging the large scale parallelism that website visitors represent. In this paper we describe our web-based distributed computing framework, Panther, and perform in-depth performance experiments for two benchmarks using real world hardware and real world browsing habits for the first time. By exploring the performance characteristics of our approach we demonstrate that this is viable for urgent workloads, but there are numerous caveats, not least the most appropriate visitor patterns to a website, that must be considered.
\end{abstract}


\section{Introduction}

The global pandemic has demonstrated the need to make urgent, accurate, decisions for complex problems. Each year there are many localised emergencies including wildfires, traffic accidents, disease, and extreme weather which, not only claim many lives and result in significant economic impact, but with the rise of global issue such as climate change, are becoming more prevalent. Modern supercomputers are designed for optimising throughput rather than the latency of individual jobs where, driven by batch queue systems, the time that a specific job will wait in the queue is unbounded. This is especially problematic for urgent workloads, such as forest fire modelling or mosquito-borne disease simulation, where there is an urgency for the job to start and deliver results. Whilst there have been some efforts to address this, such as high priority queues on HPC machines \cite{beckman2007spruce}, or federating over multiple supercomputers \cite{gibb2020bespoke} and using mathematical models to predict where to best place workloads \cite{brown2022predicting}, these do not suit every application or situation. 

An alternative is to leverage volunteer computing, where members of the public contribute their CPU processing power towards projects. The most well known approach is the Berkley Open Infrastructure for Network Computing (BOINC) \cite{anderson2004boinc} framework which specific projects such SETI@home, searching for alien life, and PrimeGrid, searching for prime numbers, have built upon. These volunteer computing projects develop applications which end users then download and install, which themselves then activate when the computer is idle by processing tasks sent to a volunteers computing from the project server.

Whilst this BOINC style approach to volunteer computing has been successful over the past twenty five years, the need for users to download and install specific software is old fashioned and arguably one of the reasons why the number of BOINC users has been slowly decreasing in recent years \cite{jones2014computer}. An obvious barriers to entry is that users must explicitly find, download and install the appropriate binary for their machine. But additionally, it is not trivial for a volunteer computing project to develop these applications in the first place, even on-top of the BOINC framework. A significant amount of time can be required in addressing cross-platform issues where projects need to develop separate binaries for each different platform type. Consequently many projects only support a subset of the overall number of platforms that BOINC is capable of running on, simply because they don't have the resources to develop for, or support, all platforms. For instance, a number of volunteer computing projects only support NVIDIA, and not AMD, GPUs due to a lack of project resources. This has become so acute that some projects require users to install virtual machines to then run BOINC under \cite{mcgilvary2013v}, which significantly raises the barrier to entry for users and can degrade performance.

In this paper we explore an alternative approach where, instead of downloading and installing software explicitly, visitors to a web page contribute some of their CPU processing power in the background. Consequently tasks comprising an urgent workloads can be matched to volunteers when they visit a website, and leveraging the web browser in this manner addresses many of the shortcomings of BOINC as well as ensuring that urgent tasks run quickly. Due to the standarisation of HTML5, urgent workload developers have a far smaller set of platforms to develop for than they would if they used BOINC. Furthermore, considering the number of visitors to some websites, there is the potential for a massive amount of parallelism. However, many of these web-sessions are short lived and, as such, a key question is whether this parallelism can be usefully leveraged. 

Whilst the major driver in this paper is the use of web-based volunteer computing for urgent workloads, there are numerous additional possibilities due to the general design of our approach. For example, it is possible for value to be generated by web visitors in ways other than advertising. For instance, with large companies that have high data processing requirements, web visitors could perform some of this processing themselves, potentially on encrypted data due to data privacy concerns, and contribute this back to the web host. The main contributions of this paper are:

\begin{itemize}
    \item The description of a web-based distributed computing framework that supports many types of application, can be installed on a wide variety of machines and leverages modern web standards.
    \item An evaluation of the performance characteristics of volunteer web-based computing in comparison to parallel C+MPI code running on real hardware for two synthetic benchmarks.
    \item An exploration of real world browsing habits and development of the concept of \emph{value} vs \emph{non-value} sessions to explain observed performance results.
    \item A comparison of WebSocket and XMLHttpRequest communication protocols to in the context of web-based distributed computing.
\end{itemize}

This paper is structured as follows, in Section 2 we describe some of the related work in urgent computing and web-based distributed computing before describing our own framework, Panther, in Section 3. In Section 4 we use our Panther framework as a basis for performance experimentation to compare, for two synthetic benchmarks, the performance one can expect from web-based distributed computing in comparison to C+MPI implementations running at the same level of parallelism on the same hardware. In this Section we then explore the impact of real world browsing habits and some of the factors that limit performance, along with the performance characteristics of different configuration options and policies. In Section 5 we draw conclusions and discuss further work.

\section{Related work}
\label{sec:relatedwork}

HPC has a long history of simulating disasters after the event, but recent technological advances have opened up the possibility of running simulations in real-time whilst disasters are unfolding. It is not only the increased computational power of modern supercomputers that unlocks such opportunities, but also the coupling with real-time data and improved technologies enabling interaction with simulations in real-time. Disasters are thankfully relatively rare, so apart from the few highly specialised disaster tracking and relief organisations such DLR-GZS, it is not realistic to have dedicated resources set aside for these, but instead to be able to make use of existing very large scale supercomputers which normally run scientific or engineering simulations.

However, HPC machines are typically optimised for throughput and not latency of individual jobs. In short, the batch queue system means that there is an unbounded time in which simulation jobs will wait in the queue, and it is entirely useless for emergency responders to be waiting for insights from an HPC simulation job that is held in the queue whilst the forest is burning. Small scale urgent workloads previously addressed this issue by relying upon high priority queues or the ability to interrupt existing simulations \cite{beckman2007spruce}, however this is not practicable for unpredictable and dynamic situations, where the amount of computing required can be large and vary significantly as time progresses \cite{gibb2019technologies}. 

The VESTEC system \cite{gibb2020bespoke} aims to address this by federating over many supercomputers and using mathematical models \cite{brown2022predicting} to match workloads to the most appropriate machine. Whilst this demonstrates promise \cite{brown2021utilising}, there are still uncertainties and these machines are not typically suited for short-term bursting of workloads. Consequently having access to some large-scale, flexible compute resource would be highly beneficial when results, however coarse-grained, are desperately needed from urgent simulations at unpredictable times.

\subsection{Volunteer computing}
Volunteer distributed computing relies on the ability to split up problems into many parallel tasks. These must be embarrassingly parallel such that they can run independently on a client machine without requiring any communication with other tasks running on other machines. Tasks tend to transform some input data, communicated from the server at the same as the task, into some output that gets sent back to the project server. This approach follows the master/worker pattern, where the server is the master and distributes tasks to the client, worker, machines which process these and send results back. Whilst the approach sounds simple, the devil is in the detail and the workers can be highly volatile, which project servers must be capable of handling.

Whilst the area of volunteer web-based computing is immature, it is not a new idea. Driven by the rapid development of web technologies, using web browsers in this manner becomes ever more attractive with the passage of time and there have been a number of previous activities around this. One of the earlier approaches was \cite{boldrin2007distributed} which was developed just after XMLHttpRequests were standardised and these are important in this context because they supported asynchronous communications between the web browser and server for the first time. However, web technology has developed significantly in the 15 years since \cite{boldrin2007distributed} and one of the major challenges that they faced was supporting multi-threading. Due to the immaturity of web technologies at the time, their calculation kernels had to explicitly break at specific points to allow other web operations, such as user interaction, to occur. This was a major weakness of their approach and since this work, WebWorkers have been standardised in HTML5, which solves this issue by supporting multiple threads for a single page.

WeevilScout, a framework for web-based distributed computing focused around Bioinformatics, was introduced in \cite{cushing2013distributed}. This was just after HTML5 had been standardised and as such the authors were able to use WebWorkers, which mitigated many of the drawbacks of \cite{boldrin2007distributed}. Whilst the work in \cite{cushing2013distributed} was more focused around clients who explicitly know they are contributing towards computation, they do make comments about \emph{parasitic computation} which is where web visitors are unaware that they are contributing CPU processing time. In this paper they compared and contrasted JavaScript performance against that of C for a range of BioInformatics algorithms and concluded that, whilst JavaScript does tend to be significantly slower, the fact that you can have an arbitrarily large amount of parallelism driven by web visitors can ameliorate much of this. However a weakness of this work was that all their experimentation assumed web sessions last until all the tasks fully complete which does not represent real-world browsing habits, where web sessions can be very short before a user navigates away. Additionally, all their experimentation was performed within a local LAN, meaning that their networking costs were not necessarily representative of the real world.

Most recently \cite{pan2017gray} considered the cost implications of web-based volunteer computing vs cloud computing. The authors developed an experimental framework heavily tied to the Amazon Web Services (AWS) technology stack and considered the unreliable nature of web page view times (known as user dwell time.) They drew on the work done in \cite{liu2010understanding} which highlighted that the vast majority of browsing habits can be expressed by a Weibull distribution \cite{rinne2008weibull} and the authors of \cite{pan2017gray} suggested a number of approaches which might optimise web-based volunteer computing in this context. However, crucially, all their experimentation was not performed on physical computing resources, but instead modelled via simulations written in MATLAB, hence their exploration of predicted results was limited. There is also a significant question around whether theory equals practice, and if their conclusions around task size and scheduling policy hold when these technologies are used by real web browsers running on real machines. Additionally, all their communications leverage XMLHttpRequests, as indeed all the related work in this section does. Communication technologies have further developed and a key question is whether the more modern WebSocket protocol which opens two-way communication streams, will provide better communication performance as it was designed to do.

The most common web-based background processing technology is currently CoinHive \cite{ruth2018digging}. This is not a general purpose volunteer computing framework, but instead used to mine the Monero crypto-currency in the web browser. CoinHive uses a centralised server and website owners are then paid based on the proportion of hashes computed by their visitors. Even though this is has a very specific focus, a number of observations can be drawn from their work. The authors choose Monero rather than other crypto-currencies, such as Bitcoin, because of the limitations of the web browser, and specifically the fact that a browser can not directly access the GPU. This is important because it illustrates that, to be successful, a framework needs to work around the limitation imposed by the web browser. They also rely on the fact that the calculation of a single hash is a very short operation, which is important due to the unreliable nature of web browser session time and the fact that browsers do not store any state between visits to a page.

\section{Panther framework}
\begin{figure}[htb]
\centering\includegraphics[width=\linewidth]{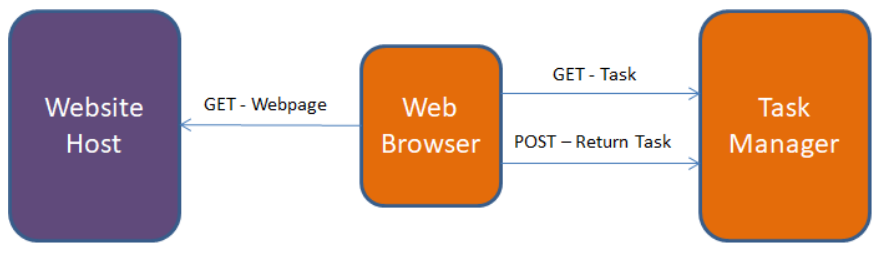}\caption{Panther framework high level architecture}

\label{fig:frameworkinteraction}
\end{figure}

We have developed Panther, a framework for volunteer web-based distributed computing whose high level architecture is illustrated in Figure \ref{fig:frameworkinteraction}. The volunteer's web browser navigates to a website on a specific host web server, however this does not host any of the tasks that the browser will process or management functionality around these. Instead, the web page served from the website host embeds a small piece of code which directs the user's web browser to communicate with Panther's task manager, running on an entirely separate server, via cross-origin HTTP requests. This approach provides two general advantages, firstly it makes it trivial for web site owners to include their visitors in the volunteer computing project because they themselves need not run the task manager which would likely affect website performance and could require site owners upgrading servers. Secondly, this architecture means that the task manager services visitors from many different websites via cross-origin HTTP requests and-so acts as a single point of truth that can be managed by a specific distributed computing project and distribute tasks with a \emph{high level view} of a large number of web browsers irrespective of the specific site that they are visiting.

\subsection{Panther's task manager}
\begin{figure}[htb]
\centering\includegraphics[width=\linewidth]{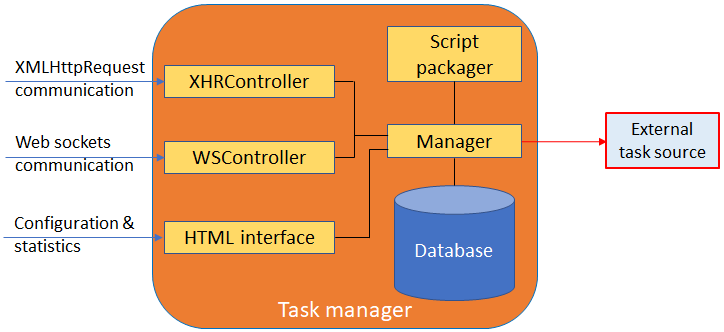}\caption{Panther task manager}

\label{fig:panthertaskmanager}
\end{figure}

The task manager of Figure \ref{fig:frameworkinteraction}, which is server side code and developed in Node.js, is the most complex aspect of our approach. Unlike \cite{pan2017gray}, our task manager is not tied to any specific vendor technology stack and can either be hosted on a physical server or the cloud as long as Node, a very common JavaScript framework, is supported. Figure \ref{fig:panthertaskmanager} illustrates the internal structure of the Panther task manager, where \emph{XHRController} and \emph{WSController} are the main entry points for user's web browser. \emph{XMLHttpRequest} \emph{WebSocket} communication protocols are provided by \emph{XHRController} and \emph{WSController} respectively. Furthermore, these modules are also responsible for data compression and decompression, and JSON parsing. We designed our approach to be plugable, so it is trivial to integrate future communication protocols as they become popular.

The \emph{manager} module is the core of the Panther task manager and periodically interacts with some \emph{external task source}. This task source can be anything that generates tasks and conforms to our API, for instance the VESTEC server which provides urgent computing tasks. Regardless, the \emph{manager} module is responsible for obtaining tasks from this task source, storing these tasks, managing the retrieval of the next task(s) for clients, monitoring the completion status of tasks, logging, and returning completed task results back to the task source.

The database of Figure \ref{fig:panthertaskmanager} provides some general persistence for logging and management. We have taken the design decision not to persist tasks however, and as such tasks only exist in the memory of the Panther task manager. This is done for performance as database look ups are comparatively slow, but does limit the number of tasks due to memory constraints and in the event of server failure then all task data on that server is lost. Whilst this might seem like a disadvantage for urgent computing workloads, the task manager is only present to provide web-based management and distribution functionality for some external task source. Consequently, as the number of tasks is dynamic when tasks complete their results are sent back back to the task source and the memory allocated to these freed up so it can be re-used by new tasks. Our design decision is that it is the responsibility of the external task source, such as the VESTEC system, to persist tasks and protect against failure.

The \emph{script packager} of Figure \ref{fig:panthertaskmanager} will automatically combine task JavaScript code with any required modules and the Panther client side API into a single, minimised, file. This combination is required because WebWorkers, which provide browser level multi-threading, can only be initiated with a single JavaScript file. The \emph{HTML interface} supports administration functionality such as configuration and viewing of server performance statistics. 

\subsection{Client side code}
\label{sec:clientsidecode}
The developer of an urgent workload must write a version of their kernel in JavaScript following a similar format to the example illustrated in Listing \ref{lst:clientcode} which adds up two numbers. There is no restriction on the code that can be written, for instance it can include further function calls, or additional modules (which are automatically packaged in the \emph{script packager} of Figure \ref{fig:panthertaskmanager}). Once the function exits the task is deemed as completed and results are sent back to Panther's task manager with the next task then executed. The \emph{task} variable argument provided to the function contains the task's data as JavaScript objects (in this example numbers \emph{a} and \emph{b}). This data can be updated by the task itself, for instance in Listing \ref{lst:clientcode} writing the result of adding numbers \emph{a} and {b} to the \emph{result}, and is automatically sent back to the task manager upon task completion.

\begin{lstlisting}[frame=lines,caption={Panther Javascript task example}, label={lst:clientcode}]
const my_task=function(task, panther) {
    task.result=task.a+task.b;
}
\end{lstlisting}

The Panther framework also provides some general API functionality that developers can use to help manage and optimise their task execution. This is accessible via the \emph{panther} argument provided to the task function, which programmers can call specific methods on. Importantly, this common interface abstracts the programmer from the underlying communication mechanism, with different class implementations provided for the XMLHttpRequest and WebSocket communication protocols. The two main API functions that a programmer might call in their task code are:

\begin{itemize}
    \item \textbf{AssignNextTask}: Retrieves the next task from the task manager, this allows programmers to explicitly request the next task(s) ahead of time whilst they are still processing an existing task.
    \item \textbf{CheckpointTask}: Writes back the state of the task to the task manager. This is useful as it allows for results to be trickled back to the Panther task manager. For instance, if the user was to navigate away from a specific web page before the task finally completes, then by checkpointing periodically not all processing is lost.
\end{itemize}

From the perspective of the visited website, the website owner need only embed a seven lines of JavaScript code on their page. Based on this, a WebWorker begins on the visitor's web browser, running in the background to obtain tasks from the Panther task manager and start to process them. 

%\begin{lstlisting}[frame=lines,caption={JavaScript code required in external website},label={lst:externalwebsitecode}]
%<script>
%  var oReq=new XMLHttpRequest();
%  oReq.addEventListener('load', function() {
%    var worker=new Worker(window.URL.createObjectURL(new Blob([this.responseText])));
%  });
%  oReq.open("get", "http://panther.address/interface");
%  oReq.send();
%</script>
%\end{lstlisting}

\subsection{Dealing with short web sessions}
\label{sec:taskscheduling}
A major challenge for web-based distributed computing is that many people visit a website for a short period of time. The way web browsers are designed means that as soon as someone navigates away from the site, then not only does processing of the tasks stop, but all context about these tasks is also lost. Consequently, there is a high likelihood that tasks will not complete and need to be rescheduled elsewhere. To counter this, the Panther framework provides several options and policies for project owners to suits their specific project. 

As described in Section \ref{sec:clientsidecode}, tasks can be checkpointed and partial results sent back to the server. The idea is that, instead of tasks being \emph{all or nothing}, results can be trickled back. The Panther task manager is capable of tracking these checkpoints and partial updates using numeric IDs, such that it will automatically update previously partial results with new ones if necessary. Internal to the task manager, a task queue is maintained and when a task has been requested by a client web browser it is removed from the head of the queue and appended to the tail. This approach handles the resending of tasks and addresses the situation of sessions ending prematurely and tasks being dropped. When a task completes it is then removed from the task queue and when results, either final or partial, arrive at the task manager, a check is performed to ensure that this task is not at a later stage by another browser. When partial results are received, the task (which will be at some point in the queue) is updated with these and as such if, and when, that task is sent out again to another web browser, only the portion of the in-complete work will be required. 

\begin{figure}[htb]
\centering\includegraphics[width=\linewidth]{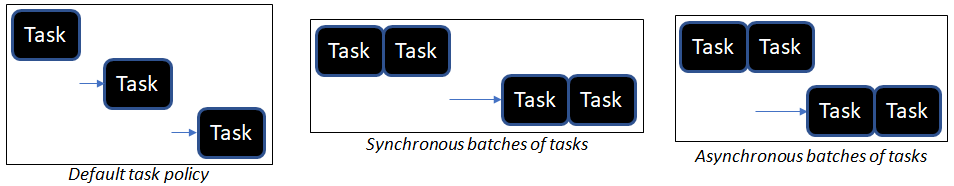}\caption{Task scheduling policies}

\label{fig:taskscheduling}
\end{figure}

Figure \ref{fig:taskscheduling} illustrates different task scheduling policies provided by the Panther framework, with the arrows indicating data transfer time. Our approach provides capabilities for maximising the useful processing time of a website visitor and obtaining as much work done by them in the, potentially short, time that they spend on the page. The default setting is a standard, synchronous, task request where the client requests a single task from the server, processes the task, sends the results back and then requests another task. Instead of sending single tasks it is possible for the Panther task manager to send batches of tasks to the client instead and on receiving these multiple tasks the Panther browser client API will store them locally ready for processing. 

It is also possible to use asynchronous communication between the web browser and Panther task manager, where browsers process a task whilst additional tasks are requested and received. The project owner can set a specific threshold value that will automatically start an asynchronous task request when the number of remaining tasks held locally reduces to this level. If the time taken to receive the new batch of tasks is less than the time taken to process the remaining task, potentially communication overhead can be eliminated. The cost of this is that a higher load placed upon the server, as browsers are requesting tasks that they might never complete due to the visitor navigating away.

\section{Evaluation}
\begin{figure}[htb]
\centering\includegraphics[width=\linewidth]{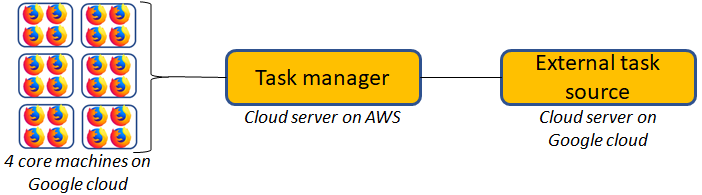}\caption{Experiment set-up}

\label{fig:experimentsetup}
\end{figure}

If web-based distributed computing is to be considered worthwhile for urgent workloads, then an important question is how the performance of our approach compares against more traditional computing, such as running code natively on volunteers machines. The set-up of our experiment hardware is illustrated in Figure \ref{fig:experimentsetup}, where we have six, four core, 15GB DRAM machines in Google cloud acting as website visitors. Each of these machines runs a separate instance of Firefox (version 67.0) on each Xeon 2.20 Ghz core (four instances of Firefox per machine.) This not only allows 24 browsers to run in parallel, but as these cores can communicate with each other it is also possible to run parallel C+MPI based codes for performance comparisons. All results reported are averaged over 5 runs.

Selenium \cite{gojare2015analysis}, a library for automated testing of web pages was used to drive the tests and this enabled Firefox to run in headless mode (no GUI) using Mozilla's GeckoDriver \cite{petre2017user} to connect the headless Firefox with Selenium. The Panther task manager was hosted on a dedicated AWS server and the external task source on a Google cloud server. It might seem strange mixing these different cloud providers, but for a fair experiment we wanted to ensure that there was a realistic communication cost (i.e. outside of a specific data centre) between the components that need to communicate in our approach.

For purposes of experimentation, two synthetic benchmarks have been adopted. Firstly the calculation of PI using the Monte Carlo method, where a circle inside a square is represented and points are picked randomly. The ratio of points inside and outside the circle can be used to calculate PI. The second benchmark is the calculation of the Mandelbrot set which involves iterating over points and testing whether each point diverges or not for a specific function. Whilst these are not urgent workloads per-se, by using benchmarks we are able to more easily identify and isolate the key performance characteristics. Furthermore, these two benchmarks test different things; the Monte Carlo benchmark is compute bound whereas the Mandelbrot benchmark is data bound as the entire local domain must be communicated as input data with the task. In all experiments, the benchmarks were prepared, configured and stored on the \emph{external task source} server ready for retrieval by the Panther task manager. By default, the Monte Carlo benchmark was split up into 720 tasks each of 200 million local iterations. The Mandelbrot benchmark used a global problem size of 400 by 300 pixels, again split over 720 tasks by default.

% FOr each benchmark describe more info - domain size etc. Where were C versions from (can we cite an external source?)

% Can re-use a core (720 tasks!)

Table \ref{tab:overallperformance} illustrates the runtime for the two benchmarks, between versions written in C (GCC 6.4) and parallelised using MPI (MPICH 3.2.1) running over the 24 cores, and versions written in JavaScript running under the web browser via our Panther framework on the same Google cloud machines. In the later case the runtime is determined by the total time required to run all the tasks and web browsers were configured such that their session would remain active for as long as the Panther task manager was sending them tasks. Based on the previous results of \cite{cushing2013distributed}, we had assumed that JavaScript versions would struggle to beat native implementations in C which isn't necessarily a problem because it is possible to gain much more parallelism via web-based distributed computing in contrast to explicitly installing and running executables. 

We were therefore surprised when the Panther Monte Carlo benchmark significantly outperformed the C+MPI version. From looking at this in more detail we found that the reason for this performance difference is that JavaScript's random number generation in Firefox is faster than the C version, not least because in C additional operations are needed to convert the number between 0 and 1. 

\begin{table}[h]
  \begin{center}
  \caption{Performance results comparing C with MPI versions running natively against those written in JavaScript using the Panther framework under the web-browser.}
  \label{tab:overallperformance}
  
  \begin{tabular}{|c|c|c|}
    \hline
    Benchmark & Configuration & Runtime (s) \\
    \hline
        \multirow{2}{*}{Monte Carlo} & C+MPI & 432\\
        \cline{3-3}
        & Panther & 156\\
    \hline
    \multirow{2}{*}{Mandlebrot} & C+MPI & 4\\
        \cline{3-3}
        & Panther & 35\\
  \hline
\end{tabular}

  \end{center}
\end{table}

The Mandelbrot benchmark results of Table \ref{tab:overallperformance} are more in-line with expectations, where the C+MPI version was nine times faster than the JavaScript version running under the Panther framework on the web browser. There are two aspects which limit the JavaScript performance here, firstly the raw computational power, as all things being equal, running natively is faster than running under the web browser. Secondly the cost of data transfer where MPI is simply much more efficient when it comes to communication than the web browser is.

\subsection{Real world browsing habits}

The experiments performed so far rely on the naive assumption that all web clients are equal and make an equal contribution towards the processing of tasks. Ultimately this assumes that all visitors stay on a website for an equal amount of time which is not realistic. The key metric here is \emph{user dwell time}, which is the time a user spends viewing a specific web page before navigating to a new page or closing the browser. A single user dwell time observation can be thought of as the total time available for that single worker to process tasks. \cite{liu2010understanding} established that the Weibull distribution \cite{rinne2008weibull} accurately models the dwell time of website visitors on individual web pages and that the vast majority (over 98\%) of visitor interaction patterns fit a Weibull distribution function with a shape parameter of 1.0 or less. This indicates a positive aging effect meaning that the longer a session remains active, the higher probably it has of remaining so. In regards to web browsing behaviour, users are more likely to leave a web page early on, for example, when quickly traversing through web pages to find interesting content to view, and more likely to remain on a page the longer they view it. 

\begin{figure}[htb]
\centering\includegraphics[width=\linewidth]{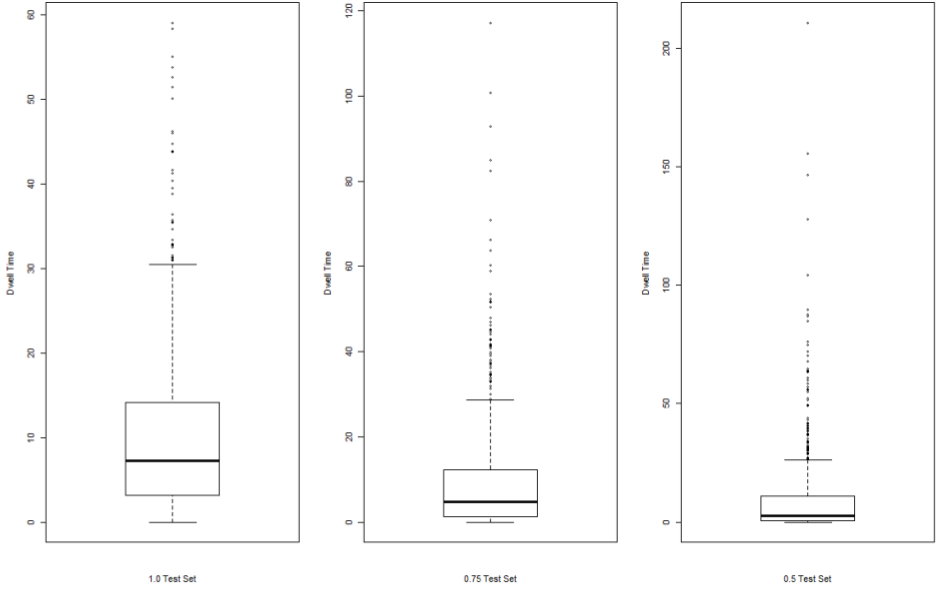}\caption{Boxplot of generated dwell times in seconds based on the Weibull distribution for different shape parameter values}

\label{fig:boxplotdweltimes}
\end{figure}

In line with the positive aging Weibull distributions, real world web page viewing habits are defined by many quick viewing sessions and fewer longer viewing sessions. The exact value for the shape parameter depends on the visitor pattern of the website in question and in this work we consider three different values, 0.5, 0.75 and 1.0 for the shape parameter. For each of these values, sets of random dwell times were generated and the boxplot distribution of these is illustrated in Figure \ref{fig:boxplotdweltimes}. For each of these shape parameters Figure \ref{fig:boxplotdweltimes} illustrates the dwell time interquatile range (the box), the medium dwell time (the black band) and outlier dwell times.

\begin{figure}[htb]
\centering\includegraphics[width=\linewidth]{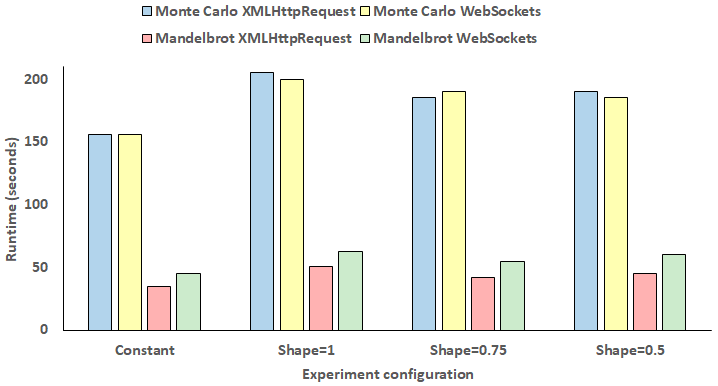}\caption{Comparison of constant runtime against that of the three different dwell time distributions for both WebSockets and XMLHttpRequest communication protocols}

\label{fig:shapesperformance}
\end{figure}

The previous experiments were re-run, but this time each browser session was allocated a specific dwell time in the generated distribution, and after the dwell was reached the browser forcibly closed (the Panther task manager automatically resends out any lost tasks). The browser then reopens on that core to represent a completely new session (i.e. a new website visitor), with a new dwell time allocated. Therefore in our experiments at any one point in time there can be as many as 24 concurrent website visitors, with many visitors over the entire run of the benchmark. The results of this experiment are illustrated in Figure \ref{fig:shapesperformance}, where \emph{constant} represents the performance observed previously with the naive assumption of each browser contributing equally, and in addition results for each different shape parameter distribution of dwell time for both benchmarks. Results using both the WebSockets and XMLHttpRequest communication protocols are reported.

It can be seen that real world browsing habits do impact performance, where the fastest completion time was 185 seconds for the Monte Carlo benchmark, at a shape of 0.75, in contrast to 156 seconds for the constant's runtime. Likewise, with the Mandelbrot benchmark, the fastest completion time was 42 seconds at a shape of 0.75 compared to the constant's 35 seconds. Both benchmarks exhibit the same pattern, where only marginal performance differences occurred between shapes 0.75 and 0.5, but both these distributions out-performed the 1.0 distribution. The results demonstrate that there is no large difference in performance between the XMLHttpRequest and WebSocket communication protocols for the Monte Carlo benchmark (max 1\%), which is compute bound, but there is a more significant difference for the Mandelbrot benchmark (max 33\%), which is data transfer bound, suggesting that the XMLHttpRequest protocol is more suited for handling larger data transfers. 

To explore the performance characteristics behind these shapes we adopt the concept of \emph{value sessions} which are where the web browser has navigated to the client web page, completed and returned at least one task before navigating away from the page. In contrast, a \emph{non-value session} is where the web browser has navigated to the client web page but not completed any tasks before navigating away from the page. It is important to understand how browsing behaviour impacts these two types of session because in the case of non-value sessions the task manager is committing resources, such as sending out the JavaScript code bundle and the task data, but gaining no return for it.

\begin{figure}[htb]
\centering\includegraphics[width=\linewidth]{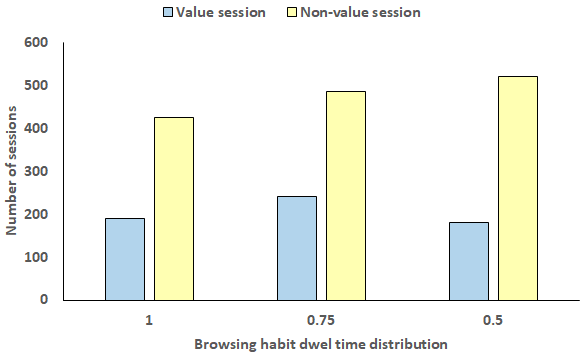}\caption{Number of value vs non-value sessions for different browsing habit dwell time shape distributions for the Monte Carlo benchmark and XMLHttpRequest protocol}

\label{fig:valuevsnonvalue}
\end{figure}

For brevity, in this paper we focus on the Monte Carlo benchmark due to the longer running time and the fact that both benchmarks exhibit the same pattern of behaviour. Figure \ref{fig:valuevsnonvalue} illustrates the number of value and non-value sessions for the three different browsing habit dwell time shape distributions for this benchmark. The best case value-session result is only 40\% of the total sessions and, as expected with the Weibull distribution, the numbers of value-session decreases as the shape value is lowered. The 0.75 and 0.5 shape parameters result in only 33\% and 25\% value sessions respectively. Under all these dwell time distributions, the majority of visits to the client web page do not result in the completion of a single task, and each one of these sessions is costing overhead on the server without any return. The average processing time of a Monte Carlo task at this granularity is 3.75 seconds, suggesting that the majority of sessions are not remaining active long enough to complete the initiation process with the Panther server and then process a task of this size in that time. The fact that, irrespective of the dwell time distribution, over half the connections with client web browsers will be useless, as far as task processing is concerned, is an consideration when selecting task granularity and undertaking performance predictions.

\begin{figure}[htb]
\centering\includegraphics[width=\linewidth]{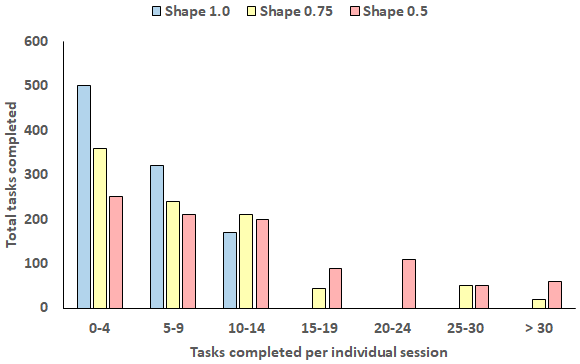}\caption{Number of tasks completed per individual session, Monte Carlo benchmark and XMLHttpRequest protocol}

\label{fig:tasksindividsession}
\end{figure}

However there is an anomaly when considering the results of Figures \ref{fig:shapesperformance} and \ref{fig:valuevsnonvalue}, where the shape 1.0 scenario has the highest amount of value sessions during the test run, but produced the slowest runtime. The distribution of the number of tasks completed by each individual session is illustrated in Figure \ref{fig:tasksindividsession}, and it can be seen that, for a shape of 1.0, no sessions completed fifteen or more tasks. However, with shapes of 0.75 and 0.5, some individual sessions complete thirty or more tasks. In-fact with a shape of 0.75, one session completed sixty four tasks, which is 6.4\% of the total task population.

Therefore that shape parameters 0.75 and 0.5 perform better than 1.0 due to the availability of longer sessions, where a larger amount of uninterrupted task processing can take place. For the shape parameter of 0.75, 64\% of the task processing was completed by only 28\% of the total value sessions, and each of these sessions completed at least five or more tasks.Based on the average task completion time, these sessions had a lifetime of at least 23 seconds. With this same shape parameter only 36\% of task processing was completed by 72\% of the value sessions, and these sessions completed four or less tasks. When we include all sessions, whether they provided any value or not, this means 64\% of the task processing was completed by only 9.3\% of all the sessions that visited the web page. Consequently, when considering what types of website traffic works best, it is far more important to focus on the traffic patterns that favour some sessions with longer dwell times as these are the \emph{superstars} that processes the majority of tasks. 

\subsection{Task granularity}

\begin{figure}[htb]
\centering\includegraphics[width=\linewidth]{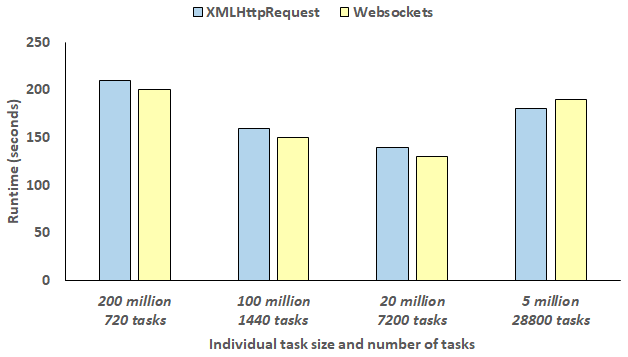}\caption{Impact of task granularity and number of tasks on the runtime for Monte Carlo benchmark and shape parameter=1.0}

\label{fig:taskgranularity}
\end{figure}

Urgent workloads can often be decomposed into a large number of tasks, and it was our hypothesis that being able to tune the granularity of tasks would be important for two reasons; firstly smaller tasks will often result in more value sessions, where at least one task completes. Secondly, smaller individual tasks will results in an increased number of total tasks, which potentially causes increased traffic to and from the Panther task manager. This is illustrated in Figure \ref{fig:taskgranularity} for the Monte Carlo benchmark which demonstrates that shrinking the task size from the default setting of 200 million local iterations and 720 tasks, initially reduces the total runtime. For instance, halving the individual task size to 100 million iterations, and doubling the number of tasks, results in an average task processing time of 1.87 seconds and a reduction in the overall run time by 11.27\%. Reducing individual task size still further to 20 million iterations, resulted in an average task processing time of 0.38 seconds and reduced the overall task run time by 15.19\% compared to the default setting of 720 tasks each of size 200 million iterations. However, there is a limit and at the smallest tested individual task size, 5 million iterations, where the average task processing time was 0.09 seconds, the run-time is only 5.39\% faster than the default.

\begin{figure}[htb]
\centering\includegraphics[width=\linewidth]{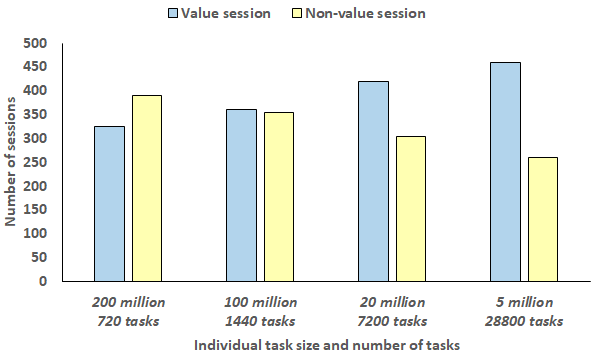}\caption{Impact of task granularity on the number of value and non-value sessions for Monte Carlo benchmark, shape parameter=1.0 and XMLHttpRequest protocol}

\label{fig:taskgranularityvaluesessions}
\end{figure}

Exploring the distribution of value against non-value sessions help to understand the impact of task granularity on runtime, and this is illustrated in Figure \ref{fig:taskgranularityvaluesessions}. It can be seen than tasks that can be processed more quickly can convert non-value sessions into value sessions. In the case of task sizes at 20 million and 5 million iterations, more value sessions are now achieved than non-value sessions. However, when comparing Figures \ref{fig:taskgranularity} and \ref{fig:taskgranularityvaluesessions}, it can be seen that the smallest task size, 5 million iterations, while achieving the most value sessions, has resulted in a slower runtime than the two larger task sizes of 20 million and 100 million iterations.

\begin{figure}[htb]
\centering\includegraphics[width=\linewidth]{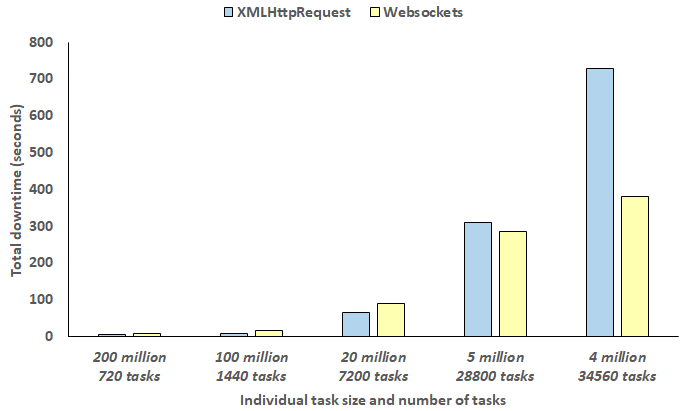}\caption{Impact of task granularity on the total downtime across all sessions for Monte Carlo benchmark and shape parameter=1.0}

\label{fig:taskgranularitydowntime}
\end{figure}

To understand why the configuration with the highest number of value sessions is not the best performing, we adopted the metric of \emph{downtime}. This is the overall amount of time that all sessions are cumulatively idle waiting for the Panther task manager to send a task for processing with no other locally buffered tasks available for the session to process. Downtime is entirely wasted time and, as such, important to minimise. Figure \ref{fig:taskgranularitydowntime} illustrates the impact of task granularity on the overall, cumulative, downtime across all sessions. It can be seen that, as the task size decreases, the downtime increases. As tasks become finer grained they also become more numerous, and hence the more interactions required with the Panther task manager. Interestingly, for 28,800 tasks at 5 million iterations we start to see a diversion between the performance of WebSocket and XMLHttpRequest communication protocols. A further experiment of 34,560 tasks each of 4 million iterations was run and the difference in downtime between the two protocols is 64.93\%. The reason for this is the lower meta-data overhead of the WebSocket communications, which is becomes more important when communication starts to dominate.

\subsection{Task scheduling policy}

\begin{figure}[htb]
\centering\includegraphics[width=\linewidth]{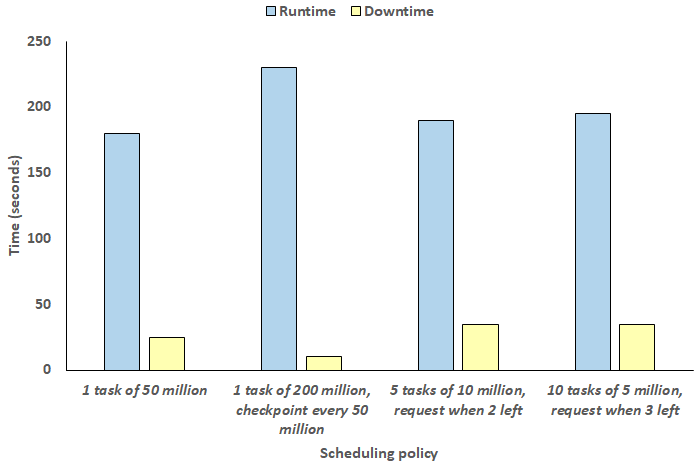}\caption{Impact of scheduling policy on runtime and downtime across all sessions for the Monte Carlo benchmark, XMLHttpRequest protocol and shape parameter=0.5}

\label{fig:schedulingpolicy}
\end{figure}

In addition to checkpointing which will trickle partial results back to the Panther task manager, there are also several scheduling policies that can be selected controlling when the next task will be sent. Figure \ref{fig:schedulingpolicy} illustrates a series of experiments using the Monte Carlo benchmark, XMLHttpRequest communication and shape parameter 0.5. The first experiment, one single task of 50 million local iterations is using the default, synchronous, scheduling policy and it is this policy that has been adopted up until this point. Consequently this configuration can be considered a base line. The next experiment uses a single task of 200 million local iterations but checkpointed, where partial results are trickled back, every 50 million iterations. At 30.5\% slower than the standard scheduling policy, this approach results in signfiicant overhead because checkpointing is synchronous and as such the task must wait until communication has completed. 

For the third and fourth experiments illustrated in Figure \ref{fig:schedulingpolicy}, multiple tasks are batched together and additional tasks are requested, asynchronously ahead of time, before all local task processing is complete. In the third experiment the web browser fetches a batch of five tasks, each of 10 million local iterations, when a client has only two local tasks remaining. In the fourth experiment batches of ten tasks, each 5 million local iterations in size, are fetched when there are only three local tasks remaining. The aim of these policies is to enable finer-grained tasks, and thus maximising the number of value sessions, whilst reducing communication overhead and downtime. However these two policies performed worse than the standard scheduling policy and this is because load on the Panther task manager was increased significantly, with the first of these batch policies increasing the load by three times and the second by five times. We do not see this same increase in load when with checkpointing however, and therefore modifying checkpointing in the future to be asynchronous will likely provide significant performance advantages.

Consequently, whilst we have provided these additional scheduling policies, for the benchmarks considered here it is beneficial to use the simplest sequential approach as it requires the least overhead in terms of communication and computational load.

\section{Ethical concerns}
In this paper we have focused on technical issues around leveraging volunteer web-based distributed computing. However there are also ethical concerns, and whilst many will agree that our objective to target urgent workloads is laudable, crucially the volunteer is no-longer the computer owner but instead the website owner. Unlike classical volunteer computing, the visitors to a website who are donating their compute time not themselves choosing to opt in or out, and furthermore as there is little transparency they are likely unaware that their machine is being used in this way. Another significant difference between our approach and classical volunteer computing is that the classical approach only undertakes computation when the computer is idle and will not impact the user experience, whereas our approach is undertaking computation whilst the computer is in use actively browsing websites.

This will not only have cost concerns, potentially increasing the power-bill for the website visitors, but furthermore there is also the danger that tasks are undertaking work that the computer owner does not morally agree with. Our approach, whilst targeted at urgent computing, is generic and could also be used with other workloads such as crypto-currency mining which visitors would likely be less enthusiastic about. 

In this work we are building upon standard HTML5 building blocks, which illustrates the importance of not visiting websites that are unfamiliar. Computer owners can undertake certain privacy actions, such as disabling JavaScript by default on sites they do not explicitly trust, and indeed this is good general practice regardless. If such web-based volunteer computing techniques become more popular, then website owners will likely need to provide visitors with an explicit choice whether to volunteer or not, potentially via an accept or reject banner which is ubiquitous on the web for handling cookies. 

\section{Conclusions}

Urgent computing on HPC machines lacks flexibility, and volunteer distributed computing is one way of providing this due to the potentially massive pool of casual compute resource that can be \emph{spun up} rapidly. Whilst classical volunteer projects, driven by user-installable frameworks such as BOINC, have seen stagnation in recent years, the enhancements made to web browsers make this technology an ideal mechanism for distributed computing. 

Leveraging the web for volunteer distributed computing is highly beneficial because there is a zero barrier to entry for end-users, and HTML5 standardisation makes developing computational kernels simpler than classical approaches such as BOINC. In this paper we have described our approach to supporting web-based volunteer computing for urgent workloads, where urgent use-case owners can download the Panther framework, run this on a variety of server types, develop JavaScript algorithms, integrate with existing urgent technologies such as VESTEC, and then deploy this with minimal changes needed to the actual websites themselves. Furthermore, our approach places minimal overhead on the website owners, requiring no additional infrastructure or support from them beyond seven lines of JavaScript embedded in their page.

We have conducted performance experiments using two synthetic benchmarks, the Monte Carlo method to find PI and calculation of the Mandelbrot set. For Monte Carlo benchmark we found that, due to the optimisation of the underlying JavaScript library, the JavaScript version of this code was significantly faster than the C+MPI version. However the Mandelbrot set benchmark was nine times slower under the Panther framework than compared against C+MPI. We then performed experimentation based on real-world browsing habits, the first time that this has been done on physical machines, and found that whilst there is some impact on the runtime this doesn't preclude the idea of web-based distributed computing. Using our concept of value vs non-value sessions, we found that the average dwell time isn't the main performance factor, but instead the number of long sessions. Inevitably these much longer sessions will be few and far between, but they contribute to task processing the most. Hence, when considering the suitability of websites based on their visitor patterns to web-based distributed computing, those that favour these few, longer dwell times, are likely to be more successful than those with larger average dwell times.

The choice between XMLHttpRequest or WebSocket communication depends on the application, XMLHttpRequest suiting kernels with fewer, larger data transfers, whereas WebSockets suits high frequency small transfers. We also considered how the granularity of tasks and scheduling policy impact performance, where results indicated that attempts to gain more value sessions from shorter dwell times result in additional communication overhead which has an impact on all viewing sessions. Most importantly, as longer viewing sessions are doing the bulk of the task processing, they are the most valuable. When optimising for shorter sessions one disadvantages the longer sessions, and the gain in the number of value sessions is not enough to make up for this loss. This contradicts conclusions drawn in \cite{pan2017gray}, but crucially those authors only simulated the runs under MATLAB, rather our use of browsers running on physical machines.

This area of web-based distributed computing is still in its infancy and not yet been deployed in the real-world for urgent workloads, and therefore we believe that there is a significant opportunity for further research and development. One obvious next step is to support kernels written in WebAssembly which should increase performance significantly.Throttling should also be considered, as currently the assumption is that a task will consume as much CPU on a single core as needed whereas in reality this is too simplistic. It will also be interesting to integrate with existing urgent technologies, such as VESTEC, enabling the consumption of results between web-based volunteers and HPC machines, potentially exploiting less accurate web-based computing results returned sooner, whilst waiting for the higher resolution HPC job to start executing.

\section*{Acknowledgement}
For the purpose of open access, the author has applied a Creative Commons Attribution (CC BY) licence to any Author Accepted Manuscript version arising from this submission.

\bibliographystyle{IEEEtran}
\bibliography{main.bib}

\end{document}